# ON BOUNDED INTEGER PROGRAMMING [*]

THÂN QUANG KHOÁT [†]

**Abstract.** We present an efficient reduction from the *Bounded integer programming* (BIP) to the *Subspace avoiding problem* (SAP) in lattice theory. The reduction has some special properties with some interesting consequences. The first is the new upper time bound for BIP, $poly(\varphi) \cdot n^{n+o(n)}$ (where $n$ and $\varphi$ are the dimension and the input size of the problem, respectively). This is the best bound up to now for BIP. The second consequence is the proof that #SAP, for some norms, is #P-hard under *semi-reductions*. It follows that the counting version of the *Generalized closest vector problem* is also #P-hard under semi-reductions. Furthermore, we also show that under some reasonable assumptions, BIP is solvable in probabilistic time $2^{O(n)}$.

**1. Introduction.** BOUNDED INTEGER PROGRAMMING is a familiar problem with many computer scientists and mathematicians. BIP asks for an integral vector $x$ satisfying the system $Ax = b$ of equations and some constraints $0 \leq x \leq u$. If there is no upper bound on the variables, it is called *Integer Programming* (IP). It is called the *Bounded Knapsack Problem* (BKP) if BIP has unique equation. If there is no upper bound on the variables in BKP, then it is called the *Knapsack Problem* (KP). These problems are extensively surveyed in the literature, [10]. However, up to now they still need much time to be solved.

In 1983, Lenstra [7] first showed that IP is solvable in polynomial time when the dimension is fixed. After this breakthrough result, researchers continue to improve it and thus many substantial improvements were proposed. The most remarkable result is from [5], where Kannan showed that IP is solvable in deterministic time $poly(\varphi) \cdot n^{2.5n}$. In his proof of this time bound, lattice problems and approximating subspaces play an important role.

Recently, Khoát [6] showed that BIP is solvable in deterministic time $poly(\varphi) \cdot n^{2n+o(n)}$. Moreover, there are some more interesting results for some other problems. For example, IP in standard form was shown to be solvable in time $poly(\varphi) \cdot n^{n+o(n)}$. He obtained these results by reducing these problems to KP, and then reducing KP to SAP, a lattice problem. The reduction from IP to KP is almost efficient in the sense that it preserves the time bound. However, the reduction from BIP to KP is inefficient. The reason is that the number of variables increases doubly after the reduction. Thus, the time bound can not be preserved.

The aim of this paper is to reduce the time bound for BIP by trying to attack it in another way. It is known that there is a reduction from BIP to BKP, which preserves the number of variables [4]. Thus, the time bound for BKP is applicable to BIP. For this reason, we are going to reduce BKP directly to SAP in lattice theory. The reduction works for both $\ell_1$ norm and $\ell_\infty$ norm, and increases the dimension to $2n + 2$, where $n$ is the dimension (number of variables) of BKP. In [6], the reduction from BKP to SAP uses KP as an intermediate problem, and increases the dimension to $4n + 2$. Consequently, the reduction here is more efficient.

---

[†]Faculty of Information Technology, Thái Nguyên University, Thái Nguyên city, Vietnam (tqkhoat@ictu.edu.vn).

The reason for our effort in reducing BKP to SAP is that SAP is efficiently reducible to the *Closest Vector Problem* (CVP) (see [8]), and that CVP can be "*efficiently*" solved. Indeed, CVP for $\ell_2$ norm is solvable in deterministic time $poly(\varphi) \cdot n^{n/2+o(n)}$ (see [3]). Furthermore, it is also the time that we need to solve CVP for $\ell_1$ norm (see Disc.1 in the Appendix). Therefore, we obtain the new time bound $poly(\varphi) \cdot n^{n+o(n)}$ for BKP (and BIP).

Another motivation of this paper is the question *whether integer programming is solvable in time* $2^{O(n)}$. This is an interesting open question. There are many lattice problems that can be solved within this time bound, and the reductions from BIP (IP) to lattice problems give an evidence for the positive answer. However, the right answer may need some complicated tools. In section 4, we study some classes of BIP that are solvable in probabilistic time $2^{O(n)}$.

We should mention more about our reduction before presenting it in detail. The reduction from BKP to SAP for $\ell_1$ norm (or $\ell_\infty$ norm) is almost *parsimonious*. Specifically, each solution to SAP corresponds to a solution to BKP; on the other hand, each solution to BKP corresponds exactly to two solutions to SAP. Therefore, #BKP may be reducible to #SAP for $\ell_1$ norm (or $\ell_\infty$ norm). Combining this with the fact that #BKP is #P-complete, we have a strong evidence that #SAP is #P-hard. The same holds for the *Generalized closest vector problem* in lattice theory. To our best knowledge, this is the first result on the hardness of #SAP and #GCVP.

The paper is organized as follows: section 2 presents the reduction from BIP to SAP and the new time bound for BIP. Some properties of the lattice of the new SAP are also investigated. Section 3 is dedicated to discussing the hardness of #SAP and #GCVP. Some special classes of BIP that can be further improved the time bound are presented in section 4.

**2. Reduction from BIP to SAP.** This section presents the main result: the reduction from BIP to SAP, and the new time bound for BIP. First of all, we recall some definitions and some conventions.

DEFINITION 2.1 (Subspace Avoiding Problem - SAP).
*Input:* a basis B of the lattice L, a subspace $M \subseteq \mathbb{R}^n$.
*Output:* the shortest vector $v \in L \backslash M$.

DEFINITION 2.2 (Closest Vector Problem - CVP).
*Input:* a basis B of the lattice L, a target vector $t \subseteq \mathbb{R}^n$.
*Output:* the vector $v \in L$ closest to t.

DEFINITION 2.3 (Generalized Closest Vector Problem - GCVP).
*Input:* a basis B of the lattice L, an affine space A in span(L), and a vector $t \in span(L)$.
*Output:* the vector $u \in L \backslash A$ closest to t.

Note that the length of a vector or the distance between two vectors mentioned in the above lattice problems is for $\ell_p$ norm, $p \geq 1$. We refer to [8] for other definitions in lattice theory.

DEFINITION 2.4 (Bounded Integer Programming - BIP).
*Find a vector $x \in \mathbb{Z}^n$ satisfying*

(2.1)
$$\begin{cases} Ax = b \\ 0 \leq x \leq u \end{cases}$$

*Where $A$ is an $m \times n$ integral matrix, $b$ is a vector in $\mathbb{Z}^m$, and $u$ is a vector in $\mathbb{Z}_+^n$.*

DEFINITION 2.5 (Bounded Knapsack Problem - BKP).
*Find a vector $x \in \mathbb{Z}^n$ satisfying*

$$(2.2) \qquad \begin{cases} a_1 x_1 + \cdots + a_n x_n = b \\ 0 \leq x \leq u \end{cases}$$

*Where the coefficients are positive integers.*

We denote by $[n]$ the set $\{1, 2, ..., n\}$, by $\|x\|_p$ the length of vector $x$ for $\ell_p$ norm ($p \geq 1$), and by $\|x\|$ the length of vector $x$ for $\ell_\infty$ norm. If there is no more explanation, $Bx$ denotes the product of a matrix $B$ with a column vector $x$. If $x$ and $u$ are vectors in $\mathbb{R}^n$, then $\frac{x}{u}$ denotes the vector $\left(\frac{x_1}{u_1}, \frac{x_2}{u_2}, ..., \frac{x_n}{u_n}\right)$.

It is easy to see that BIP is reducible to BKP (e.g. we can use Kannan's technique in [4] to aggregate the equations in BIP). Moreover, the reduction from BIP to BKP preserves the dimension (number of variables). Thus, we may say that they have the same properties.

THEOREM 2.6. *There exists a deterministic polynomial time reduction from BKP to SAP for $\ell_\infty$ norm. Furthermore, the reduction increases the dimension by $n + 2$, where $n$ is the dimension of BKP.*

THEOREM 2.7. *There exists a deterministic polynomial time reduction from BKP to SAP for $\ell_1$ norm. Furthermore, the reduction increases the dimension by $n + 2$, where $n$ is the dimension of BKP.*

COROLLARY 2.8. *There exists a deterministic polynomial time reduction from BIP to SAP for $\ell_1$ norm. Moreover, the reduction increases the dimension by $n + 2$, where $n$ is the dimension of BIP.*

Combining this fact with the results in [8], BIP is reducible to CVP for $\ell_1$ norm. Moreover, the reduction increases the dimension to $2n + 2$. From the fact that CVP for $\ell_1$ norm is solvable in deterministic time $poly(\psi) \cdot n^{n/2+o(n)}$ (see Disc 1 in the Appendix), we have the following result.

THEOREM 2.9. *BIP is solvable in deterministic time $poly(\varphi) \cdot n^{n+o(n)}$, where $n$ and $\varphi$ are the dimension and the size of the input, respectively.*

Now we present the reduction from BKP to SAP. The argument is similar to the one in [6]. Assume that we are given BKP (2.2). Without loss of generality, we assume that $a_i < b$ for all $i \in [n]$. Moreover, we assume that $\sum_{j=1}^n a_j u_j \neq 2b$. This assumption can be guaranteed by making some simple modifications to (2.2) or by reducing it to new BKP satisfying the assumption, e.g. see Disc 2 in the Appendix. Then we will deal with the subspace $S = \{x \in \mathbb{R}^{2n+2} : a_1 u_1 x_1 + \cdots + a_n u_n x_n = 0\}$. Consider the lattice $L_0$ generated by the basis $B_0$, where

$$B_0 = \begin{pmatrix} \hat{U}_n & 0 & 0 & 0 \\ s_0 a & -s_0 b & 0 & 0 \\ 0 & 0 & \hat{U}_n & 0 \\ s_1 C & s_1 \delta \lambda^n & s_1 C & -s_1 \gamma \end{pmatrix}$$

is a $(2n+2) \times (2n+2)$ matrix; $\hat{U}_n = diag(\frac{1}{u_1}, \frac{1}{u_2}, ..., \frac{1}{u_n})$ is a diagonal matrix of size $n$; $a = (a_1, ..., a_n), \delta = u_1...u_n, \delta_i = \delta/u_i, \forall i, u_{max} = \max\{u_1, ..., u_n\}$; $s_0, s_1$ and $\lambda$ are optional integers satisfying $s_0 > n^2, s_1 > n^2, \lambda > \delta n^3 u_{max}$; $C = (\delta_1, \delta_2 \lambda, ..., \delta_n \lambda^{n-1})$, and $\gamma = 1 + \lambda + \cdots + \lambda^n$.

First and foremost, some special properties of $L_0$ should be presented in order to prove the above theorems and to survey the hardness of some counting problems in the next section.

LEMMA 2.10. *For any vector $y = B_0 z \in L_0$, if $y$ does not satisfy one of the following conditions*

$$y_{n+1} = y_{2n+2} = 0 \tag{2.3}$$

$$y_i + y_{n+1+i} = z_{n+1}, \ \forall i \in [n] \tag{2.4}$$

*then $y$ has at least one coordinate with magnitude not less than $\Theta(p)$, where $p = \min\{s_0, s_1, \lambda/(\delta n u_{\max})\}$.*

*Proof.* We rewrite the vector $y \in L_0$ in more detail: for all $i \in [n]$

$$y_i = \frac{z_i}{u_i} \tag{2.5}$$

$$y_{n+1} = s_0 \left( \sum_{j=1}^{n} a_j z_j - b z_{n+1} \right) \tag{2.6}$$

$$y_{n+1+i} = \frac{z_{n+1+i}}{u_i} \tag{2.7}$$

$$\begin{aligned}
y_{2n+2} &= s_1 \left( \sum_{j=1}^{n} \delta_j \lambda^{j-1} z_j + \sum_{j=1}^{n} \delta_j \lambda^{j-1} z_{n+1+j} + \delta \lambda^n z_{n+1} - \gamma z_{2n+2} \right) \\
&= s_1 \left( \sum_{j=1}^{n} \delta_j \lambda^{j-1} (z_j + z_{n+1+j}) + \delta \lambda^n z_{n+1} - \gamma z_{2n+2} \right)
\end{aligned} \tag{2.8}$$

It is easy to see that if $y_{n+1} \neq 0$ then $y_{n+1}$ is a nonzero multiple of $s_0$. This means $|y_{n+1}| \geq s_0 \geq p$. The same holds for $y_{2n+2}$.

Now assume that $y$ does not satisfy (2.4), but (2.3). Then there exists $i \in [n]$ such that $y_i + y_{n+1+i} \neq z_{n+1}$. From (2.5) and (2.7), vector $z$ holds the same property, i.e. $z_i + z_{n+1+i} \neq u_i z_{n+1}$ or $\delta_i(z_i + z_{n+1+i}) \neq \delta z_{n+1}$. By the assumption $y_{n+1} = y_{2n+2} = 0$, we have

$$\sum_{j=1}^{n} a_j z_j = b z_{n+1} \tag{2.9}$$

$$\sum_{j=1}^{n} \delta_j \lambda^{j-1}(z_j + z_{n+1+j}) + \delta \lambda^n z_{n+1} = \gamma z_{2n+2} \tag{2.10}$$

Let $w = (w_1, ..., w_{n+1}) \in \mathbb{Z}^{n+1}$ such that $w_i = \delta_i(z_i + z_{n+1+i}), \forall i \in [n]$, and $w_{n+1} = \delta z_{n+1}$. Equation (2.10) can be rewritten by:

$$\sum_{j=1}^{n} \lambda^{j-1} w_j + \lambda^n w_{n+1} = \gamma z_{2n+2} \tag{2.11}$$

We see that there exists $i$, $w_i \neq w_{n+1}$ (due to the known property of $z$). That is, $w$ is an integral solution to (2.11) other than $(z_{2n+2}, ..., z_{2n+2})$. By Lemma 4.7 in the appendix, $w$ has at least one component whose absolute value is not less than $\Theta(\lambda)$. From this fact, we can finish the proof by examining the following cases:

*Case 1:* $|w_{n+1}| \geq \Theta(\lambda)$

We immediately have $|\delta z_{n+1}| \geq \Theta(\lambda)$; thus, $|z_{n+1}| \geq \Theta(\lambda/\delta)$. From the fact that $a_i < b, \forall i \in [n]$, every integral solution to (2.9) has at least one component $z_h(h \in [n])$ such that $|z_h| \geq \Theta(\lambda/\delta n)$. This implies $|y_h| \geq \Theta(\lambda/\delta n u_i) \geq \Theta(p)$. The desired property of $y$ follows.

*Case 2:* $|w_i| \geq \Theta(\lambda)$, for some $i \in [n]$

This case leads to $\delta |z_i + z_{n+1+i}| \geq \delta_i |z_i + z_{n+1+i}| \geq \Theta(\lambda)$. It is not hard to see that either $|z_i| \geq \Theta(\lambda/\delta)$ or $|z_{n+1+i}| \geq \Theta(\lambda/\delta)$. From (2.5) and (2.7), we have either $|y_i| \geq \Theta(\lambda/\delta u_i)$ or $|y_{n+1+i}| \geq \Theta(\lambda/\delta u_i)$. The proof ends. □

LEMMA 2.11. *Suppose that (2.2) has solutions. For every $y^* \in L_0 \backslash S$, either $(u_1 y_1^*, ..., u_n y_n^*)$ or $-(u_1 y_1^*, ..., u_n y_n^*)$ is a solution to (2.2) if $\|y^*\| \leq 1$.*

*Proof.* From lemma 2.10, $y^* = B_0 z$ satisfies (2.3) and (2.4). Then we immediately have

$$\sum_{j=1}^{n} a_j z_j = b z_{n+1} \tag{2.12}$$

$$y_i^* + y_{n+1+i}^* = z_{n+1}, \forall i \in [n] \tag{2.13}$$

From the hypothesis $y^* \in L_0 \backslash S$, $y^*$ cannot be in $S$. That is, $\sum_{j=1}^{n} a_j u_j y_j^* \neq 0$; thus $\sum_{j=1}^{n} a_j z_j \neq 0$. Combining this fact with (2.12) yields $z_{n+1} \neq 0$. From (2.13), it is not hard to see that if $|z_{n+1}| > 2$, then either $|y_i^*| > 1$ or $|y_{n+1+i}^*| > 1$. This means $\|y^*\| > 1$, contrarily. Consequently, the remainder of the proof is to consider the case $0 < |z_{n+1}| \leq 2$.

Assume that $z_{n+1} = 2$. (The case $z_{n+1} = -2$ can be dealt with similarly.) Then we have $y_i^* + y_{n+1+i}^* = 2$, $\forall i \in [n]$. An immediate observation is that if $y_i^* = y_{n+1+i}^* = 1$, $\forall i \in [n]$, then $z_i = z_{n+1+i} = u_i$, $\forall i \in [n]$. From (2.12) we have $\sum_{j=1}^{n} a_j u_j = 2b$, being contrary to the assumption $\sum_{j=1}^{n} a_j u_j \neq 2b$. Hence, there exists $r \in [n]$ such that either $|y_r^*| > 1$ or $|y_{n+1+r}^*| > 1$. This leads to $\|y^*\| > 1$, contrarily.

If $z_{n+1} = 1$, then we have $\sum_{j=1}^{n} a_j z_j = b$ and $y_i^* + y_{n+1+i}^* = (z_i + z_{n+1+i})/u_i = 1$, $\forall i \in [n]$. That is, $\sum_{j=1}^{n} a_j z_j = b$ and $z_i + z_{n+1+i} = u_i$, $\forall i \in [n]$. From this fact, the first remark is that if there exists $r \in [n]$ such that $z_r < 0$, then $z_{n+1+r} > u_r$, and thus $y_{n+1+r}^* > 1$. This means $\|y^*\| > 1$, contrarily. Another remark is that if there exists $r \in [n]$ such that $z_r > u_r$, then $y_r^* > 1$. This means $\|y^*\| > 1$, contrarily. Consequently, $0 \leq z_i \leq u_i$, $\forall i \in [n]$. That is, $(z_1, z_2, ..., z_n)$ is a solution to (2.2). Equivalently, $(u_1 y_1^*, ..., u_n y_n^*)$ is a solution to (2.2).

By the same argument we can show that if $z_{n+1} = -1$, then $-(u_1 y_1^*, ..., u_n y_n^*)$ is a solution to (2.2). The lemma follows. □

From this property, we remark that the shortest vectors in $L_0 \backslash S$ for $\ell_\infty$ norm should have length not greater than 1. Indeed, a careful observation reveals that if $\hat{x}$ is a solution to (2.2), then $\hat{y} = (\frac{\hat{x}}{u}, 0, e - \frac{\hat{x}}{u}, 0)$ is a vector in $L_0 \backslash S$, and $\|\hat{y}\| \leq 1$.[1] Therefore, $\|y^*\| \leq \|\hat{y}\| \leq 1$ if $y^*$ is the shortest vector in $L_0 \backslash S$ for $\ell_\infty$ norm. These facts quickly leads to the following corollary.

---

[1] Here $e = (1, 1, ..., 1) \in \mathbb{R}^n$.

COROLLARY 2.12. *Suppose that (2.2) has solutions. If $y^*$ is the shortest vector in $L_0 \backslash S$ for $\ell_\infty$ norm, then either $(u_1 y_1^*, ..., u_n y_n^*)$ or $-(u_1 y_1^*, ..., u_n y_n^*)$ is a solution to (2.2).*

From these observations, the claim of theorem 2.6 is easily proven.

*Proof.* [Proof of theorem 2.6] Notice that Corollary 2.12 implies a reduction from BKP to SAP for $\ell_\infty$ norm. Moreover, any parameters including $s_0, s_1, \lambda$ can be chosen in time polynomial. Thus, the reduction is a Karp one. The dimension of the lattice $L_0$ is $2n + 2$. Hence, it is also the dimension of the new SAP. Consequently, theorem 2.6 follows. ☐

We now return to Theorem 2.7. To prove it, we need some other observations about $L_0$. From the arguments in the proof of Lemma 2.11, we remark that $\|y^*\|_1 = n$. Indeed, if $z_{n+1} = 1$ then $y_j^* \geq 0, \forall j, y_{n+1}^* = y_{2n+2}^* = 0, y_i^* + y_{n+1+i}^* = 1, \forall i \in [n]$, and thus $\|y^*\|_1 = \sum_{j=1}^{2n+2} |y_j^*| = \sum_{j=1}^{2n+2} y_j^* = n$. If $z_{n+1} = -1$, then $y_j^* \leq 0, \forall j, y_{n+1}^* = y_{2n+2}^* = 0, y_i^* + y_{n+1+i}^* = -1, \forall i \in [n]$, and thus $\|y^*\|_1 = n$. In short, $\|y^*\|_1 = n$ if $\|y^*\| \leq 1$. This observation leads to the hope that if a lattice vector $y \in L_0 \backslash S$ has length not greater than $n$ for $\ell_1$ norm, then it gives a solution to (2.2). Fortunately, it is the case.

LEMMA 2.13. *Suppose that (2.2) has solutions. Then a lattice vector $y^*$ is the shortest in $L_0 \backslash S$ for $\ell_1$ norm if and only if $\|y^*\|_1 = n$. Moreover, if $y^*$ is the shortest vector in $L_0 \backslash S$ for $\ell_1$ norm, then either $(u_1 y_1^*, ..., u_n y_n^*)$ or $-(u_1 y_1^*, ..., u_n y_n^*)$ is a solution to (2.2).*

*Proof.* As mentioned earlier, $\hat{y} = (\frac{\hat{x}}{u}, 0, e - \frac{\hat{x}}{u}, 0)$ is a short vector in $L_0 \backslash S$ if $\hat{x}$ is a solution to (2.2). Moreover, $\|\hat{y}\|_1 = n$. Therefore, a short vector in $L_0 \backslash S$ should have length not much larger than $n$.

Assume that $y^*$ is the shortest vector in $L_0 \backslash S$ for $\ell_1$ norm. Then $\|y^*\|_1 \leq \|\hat{y}\|_1 = n$. Combining this with Lemma 2.10, $y^* = B_0 z$ satisfies both (2.3) and (2.4). Thus, by the hypothesis $y^* \notin S$, we have $\sum_{i=1}^{2n+2} y_i^* = \sum_{i=1}^{n} y_i^* + \sum_{i=1}^{n} y_{n+1+i}^* = n z_{n+1}$, where $z_{n+1} \neq 0$. This means $\|y^*\|_1 \geq n|z_{n+1}| \geq n$. Hence, $\|y^*\|_1 = n$.

If $\|y^*\|_1 = n$, then $y^*$ must be the shortest vector in $L_0 \backslash S$. Indeed, for every $y^1 = B_0 z^1 \in L_0 \backslash S$, if $y^1$ does not satisfy either (2.3) or (2.4), then $\|y^1\|_1 \geq \Theta(p) \geq n$ (where $p = \min\{s_0, s_1, \lambda/(\delta n u_{\max})\}$). Otherwise, we have $y_i^1 + y_{n+1+i}^1 = z_{n+1}^1, \forall i \in [n]$, where $z_{n+1}^1 \neq 0$. This means $\|y^1\|_1 \geq |\sum_{i=1}^{2n+2} y_i^1| = n|z_{n+1}^1| \geq n$. In short, $\|y^1\|_1 \geq n$ for every $y^1 \in L_0 \backslash S$. As a result, $\|y^*\|_1 \leq \|y\|_1$ for all $y \in L_0 \backslash S$. That is, $y^*$ is the shortest vector in $L_0 \backslash S$ for $\ell_1$ norm.

Suppose that $y^*$ is the shortest vector in $L_0 \backslash S$ for $\ell_1$ norm. By the above observations, we have $\|y^*\|_1 = n$. This means $|z_{n+1}| = 1$. By the same arguments as (in the last two paragraphs) in the proof of Lemma 2.11, the last statement of this lemma follows. ☐

*Proof.* [Proof of theorem 2.7] Notice that Lemma 2.13 implies a reduction from BKP to SAP for $\ell_1$ norm. Moreover, any parameters including $s_0, s_1, \lambda$ can be chosen in time polynomial. Thus, the reduction is a Karp one. The dimension of the lattice $L_0$ is $2n + 2$. Hence, it is also the dimension of the new SAP. Consequently, theorem 2.7 follows. ☐

**Remark.** We have the new time bound for BIP. Nonetheless, by binary technique, we can easily show that the optimization version of BIP has the same time bound. More concretely, consider the problem of finding a vector $x \in \mathbb{Z}^n$ such that it minimizes $c \cdot x$, subject to the constraints $Ax = b$ and $0 \leq x \leq u$. Since the upper bound and lower bound of the objective function is easily obtained, we can use binary technique to reduce this problem to BIP. Note that reduction increases the number of variables by 1. Thus, combining this facts with Theorem 2.9, we have the time bound $n^{n+o(n)}$ for the optimization version of BIP.

**3. Hardness of #SAP and #GCVP.** This section is dedicated to presenting a survey of the hardness of the problem of counting the solutions to SAP for $\ell_p$ norm. This problem is denoted by $\#SAP_p$. We also consider the problem of determining the number of solutions to GCVP for $\ell_p$ norm, denoted by $\#GCVP_p$. Since SAP is a special case of GCVP, the hardness of $\#SAP_p$ implies the one of $\#GCVP_p$.

Let us see more about the reduction in the previous section. Each short vector of $L_0\backslash S$ corresponds to a solution to (2.2) provided that (2.2) has solutions.[2] Conversely, assuming $\hat{x}$ is a solution to (2.2), $L_0\backslash S$ must contain two short vectors corresponding to $\hat{x}$ in the sense that the head parts of these two vectors are similar to either $\frac{\hat{x}}{u}$ or $-\frac{\hat{x}}{u}$. Moreover, every other short vector in $L_0\backslash S$ cannot correspond to $\hat{x}$ in the same sense. These observations reveal that if every solution to (2.2) lies on the surface of the hypercube $[0, u] = \{x \in \mathbb{R}^n : 0 \leq x \leq u\}$,[3] then counting the solutions to (2.2) is reducible to the one to SAP for $\ell_\infty$ norm as long as (2.2) has solutions.

LEMMA 3.1. *#BKP under the assumption that BKP has solutions is reducible to* $\#SAP_\infty$.

To prove this lemma, we need the following observation.

LEMMA 3.2. *If every solution to (2.2) lies on the surface of the hypercube $[0, u]$, then the $\ell_\infty$ length of the shortest vector in $L_0\backslash S$ is 1.*

*Proof.* Assume that $y^* = B_0 z$ is the shortest vector in $L_0\backslash S$ for $\ell_\infty$ norm. The arguments in the proof of Lemma 2.11 shows that $y_i^* + y_{n+1+i}^* = z_{n+1}$, where $|z_{n+1}| = 1$. By the hypothesis of $y^*$, there exists $x^*$ being a solution to (2.2) such that $x^* = (u_1^*|y_1^*|, ..., u_n^*|y_n^*|)$. Note that $x^*$ lies on the surface of $[0, u]$; that is, $\exists x_j^* \in \{0, u_j\}$. This means there exists $|y_j^*| \in \{0, 1\}$; therefore, $|y_{n+1+j}^*| \in \{1, 0\}$. Combining this with Lemma 2.11 yields $\|y^*\| = 1$. □

*Proof.* [Proof of Lemma 3.1] It is easy to see that some solutions to BKP (2.2) may not satisfy the assumption of Lemma 3.2. Thus counting the solutions to SAP for $\ell_\infty$ norm may be useless. However, some modifications to the original BKP may be worth.

Assume that BKP (2.2) does not satisfy the assumption of Lemma 3.2. We reduce it to the new BKP as described in Disc.2. The new BKP has the property that every solution to it has the last component's value being 1. That means the new BKP satisfies the assumption of Lemma 3.2. Therefore, applying the reduction from the new BKP to SAP yields the fact that every shortest vector in $L_0\backslash S$ for $\ell_\infty$ norm is of length 1, and that each solution to the new BKP corresponds exactly to two solutions to SAP for $\ell_\infty$ norm. Consequently, the number of solutions to the new BKP would be revealed if one know the solution to $\#SAP_\infty$. The proof completes. □

Note that Lemma 3.1 does not imply a reduction from #BKP to $\#SAP_\infty$. The reason is that the solution to $\#SAP_\infty$ is always a non-zero number, whereas the one to #BKP may be zero. This is caused by some disadvantages of our reduction in the previous section.

DEFINITION 3.3. *Let #A and #B be counting problems. Then #A is said to be* semi-reducible *to #B if the following satisfies. Assuming that the solution to #A is non-zero, solving #A needs polynomial calls to #B oracle.*

---

[2] Here, by saying short vector we mean that the $\ell_\infty$ length of the vector is not greater than 1.

[3] The vector $x$ is said to lie on the surface of the hypercube $[0, u]$ if there exists $i \in [n]$ such that $x_i \in \{0, u_i\}$.

*If #A is semi-reducible to #B, we say that there is a* semi-reduction *from #A to #B.*

This kind of reduction is quite weak since it only maps a counting problem with non-zero solution to another counting problem. The stronger kind of reduction for counting problems may be found in [9].

Turn to our problem, #SAP$_\infty$. Lemma 3.1 clearly implies the following result.

COROLLARY 3.4. *#BKP is semi-reducible to #SAP$_\infty$.*

It is well-known that BKP is NP-hard. The reduction from SAT to KP in [9], via numerous problems, can be made to be parsimonious. This means that #SAT is reducible to #KP. Note that #SAT is #P-complete (see Theorem 18.1 in [9]). Hence #KP is also #P-complete. Since KP is a special case of BKP, we conclude that #BKP is #P-complete. As a result, we have

THEOREM 3.5. *#SAP$_\infty$ is #P-hard under semi-reductions.*

Although this theorem does not imply the #P-hardness of #SAP$_\infty$, but it gives a strong evidence for the hardness of #SAP$_\infty$. We believe that #SAP$_\infty$ is #P-hard. Nonetheless, with this technique we cannot prove that.

By the same arguments, we can prove the following result for #SAP$_1$.

THEOREM 3.6. *#SAP$_1$ is #P-hard under semi-reductions.*

It is not hard to see that GCVP is SAP if the input affine space $A$ is a subspace, and $t = 0$. Hence SAP is a special case of GCVP. By Theorem 3.5 and 3.6, we obtain the same results for GCVP.

THEOREM 3.7. *#GCVP$_1$ and #GCVP$_\infty$ are #P-hard under semi-reductions.*

**4. BIP under some assumptions.** We continue to study some classes of BIP for which there exist algorithms running in probabilistic time $2^{O(n)}$. It is well-known that BIP is NP-hard; thus, it is likely that there is no polynomial time algorithm for BIP. However, real-life applications expect lower time bound for BIP. This demand motivates us to find more efficient algorithms for it. We believe that BIP is solvable in time $2^{O(n)}$. Nonetheless, here we are only able to show this property for some classes of BIP.

In section 2, we know that BKP is reducible to SAP for $\ell_\infty$ norm. Thus, the time bound for the new SAP is applicable to the original BKP. It is clear that there are many algorithms for SAP (see [1], [2], [8]); however, the time bound for general SAP is quite large. Fortunately, SAP can be solved in time $2^{O(n)}$ if it has some special properties, as shown in [2].

THEOREM 4.1. *(Blömer and Naewe). Let $L$ be a lattice and $M$ be a subspace of span$(L)$. Assume that there exist absolute constants $c, \varepsilon$ such that the number of $v \in L \backslash M$ satisfying $\|v\| \leq (1+\varepsilon)\lambda_M(L)$ is bounded by $2^{cn}$, where $\lambda_M(L)$ is the length of the shortest vector in $L \backslash M$. Then there exists an algorithm that solves SAP with probability exponentially close to 1. The running time is $2^{O(n)}$.*

This theorem shows that if the number of short lattice vectors in $L \backslash M$ is not too large, we may solve SAP in time $2^{O(n)}$. On the other hand, if the exact solution to SAP is not necessary in some cases, we can find an approximate solution to SAP in almost same time.

THEOREM 4.2. *(Blömer and Naewe). There exists a randomized algorithm that solves SAP with approximation factor $1 + \varepsilon$, $0 < \varepsilon \leq 2$, with probability exponentially close to 1. The running time is $((2 + 1/\varepsilon)^n.b)^{O(1)}$, where $b$ is the input size of the problem.*

We know that the new SAP resulted from our reduction may not satisfy the assumptions in Theorem 4.1, and that the approximate solutions to the new SAP may not give any solution to the original BKP. Consequently, we cannot expect to improve the time bound for BKP. However, if the original BKP has some special properties, then it can be "*efficiently*" solved.

THEOREM 4.3. *Assume that BKP (2.2) has a solution $\hat{x}$ such that $\frac{\hat{x}_i}{u_i} \in [\frac{\varepsilon}{1+\varepsilon}, \frac{1}{1+\varepsilon}], \forall i \in [n]$, for some absolute constant $\varepsilon \in (0, 1]$. Then there exists a randomized algorithm that solves BKP with probability exponentially close to 1. The running time of the algorithm is $poly(\varphi) \cdot 2^{O(n)}$, where $\varphi$ is the size of the input.*

This theorem means that BKP is solvable in probabilistic time $poly(\varphi) \cdot 2^{O(n)}$ if it has a certain solution lying rather closely to the center of the hypercube $[0, u]$. The same holds for BIP.

COROLLARY 4.4. *Assume that BIP (2.1) has a solution $\hat{x}$ such that $\frac{\hat{x}_i}{u_i} \in [\frac{\varepsilon}{1+\varepsilon}, \frac{1}{1+\varepsilon}], \forall i \in [n]$, for some absolute constant $\varepsilon \in (0, 1]$. Then there exists a randomized algorithm that solves BIP with probability exponentially close to 1. The running time of the algorithm is $poly(\varphi) \cdot 2^{O(n)}$, where $\varphi$ is the size of the input.*

*Proof.* [Proof of Theorem 4.3]

It is not hard to see that if $u_{\max}$ is small, e.g. $O(1)$, then we can solve BKP by enumerating all possible vectors in $[0, u] \cap \mathbb{Z}^n$. The time we need to do this is $O(u_{\max}^n) = 2^{O(n)}$. Thus, without loss of generality, we assume that $u_{\max}$ is not too small. Then the following is straightforward.

$$(4.1) \quad 1 - \frac{u_{\max} + 1}{u_{\max}(1 + \varepsilon)} < \frac{\varepsilon}{1 + \varepsilon}$$

On the other hand, due to $1 < 1 + 1/u_{\max}$, we have

$$(4.2) \quad \frac{1}{1 + \varepsilon} < \frac{u_{\max} + 1}{u_{\max}(1 + \varepsilon)}$$

.

If we combine the assumption of $\hat{x}$ with (4.1) and (4.2), then

$$(4.3) \quad 1 - \frac{u_{\max} + 1}{u_{\max}(1 + \varepsilon)} < \frac{\hat{x}_i}{u_i} < \frac{u_{\max} + 1}{u_{\max}(1 + \varepsilon)}, \ \forall i \in [n]$$

Applying the reduction from BKP to SAP as in section 2, we obtain the new SAP. Let the new SAP be the centre of our concentration from now on.

Notice that the vector $\hat{y} = \left(\frac{\hat{x}}{u}, 0, e - \frac{\hat{x}}{u}, 0\right)$ is in $L_0 \backslash S$, and that $\|\hat{y}\| = \max\{\|\frac{\hat{x}}{u}\|, \|e - \frac{\hat{x}}{u}\|\}$. Combining this with (4.3) yields

$$(4.4) \quad \|\hat{y}\| < \frac{u_{\max} + 1}{u_{\max}(1 + \varepsilon)}$$

.

Consider any vector $y = B_0 z$ in $L_0 \backslash S$ such that $(u_1|y_1|, ..., u_n|y_n|)$ is not a solution to (2.2). From Lemma 2.11, there exists $r$ such that $|y_r| > 1$. If $y$ does not satisfies either (2.3) or (2.4), by Lemma 2.10 we have $\|y\| \geq 2 > 1 + 1/u_{\max}$. Otherwise, $r \neq n + 1$ and $r \neq 2n + 2$. This means $|y_r| = |z_r/u_{r^*}| > 1$;[4] equivalently, $|z_r| > u_{r^*}$. From the fact that $z_r \in \mathbb{Z}$, we immediately have $|z_r| \geq u_{r^*} + 1$. These observations yield

$$|y_r| \geq \frac{u_{r^*} + 1}{u_{r^*}} \geq 1 + \frac{1}{u_{r^*}} \geq 1 + \frac{1}{u_{\max}}.$$

---

[4] Here, $r^* = r$ if $r \leq n$; otherwise, $r^* = r \mod (n + 1)$.

Therefore,

$$(4.5) \qquad \|y\| \geq 1 + \frac{1}{u_{\max}}$$

Combining (4.4) with (4.5) leads to $\|\hat{y}\| < \|y\|/(1+\varepsilon)$; equivalently, $\|y\| > (1+\varepsilon)\|\hat{y}\|$.

Let $y^*$ be a solution to the new SAP (the shortest vector in $L_0\backslash S$). Then $\|y^*\| \leq \|\hat{y}\|$. That is, $\|y\| > (1+\varepsilon)\|y^*\|$ for all $y \in L_0\backslash S$ such that $(u_1|y_1|,...,u_n|y_n|)$ is not a solution to (2.2). This implies that an approximate solution to the new SAP within the factor $1+\varepsilon$ is sufficient to give a solution to (2.2). In other words, BKP (2.2) is reducible to the problem of approximating SAP within a factor of $1+\varepsilon$.

If we know $\varepsilon$, then the new SAP can be approximated by SAP solver in [2] to find a solution to (2.2), and the running time is $poly(\varphi) \cdot 2^{O(n)}$ (due to Theorem 4.2). Conversely, we can find a solution to (2.2) as follows: repeat approximating the new SAP (by SAP solver in [2]) correspondingly with approximation factors $2, 1+\frac{1}{2}, 1+\frac{1}{4},...$ until we find a solution to (2.2) from approximate solutions of the new SAP. Since any approximate solution to the new SAP within the factor $1+\varepsilon$ is sufficient to give a solution to the original BKP, the number of repeat loops that we need to solve (2.2) is at most $\log_2(1/\varepsilon)$. Therefore, the total running time is $\hat{T} = T.\log_2(1/\varepsilon)$, where $T$ is the running time of each repeat loop. Note that $\varepsilon$ is a certain absolute constant. Consequently, we have $\hat{T} = O(T)$. The claim of the theorem is clear. □

Another property of BKP also allows the possibility of solving it in time $2^{O(n)}$.

THEOREM 4.5. *Assume that there exist absolute constant $c, \varepsilon$ such that the number of integral vectors $\hat{x}$ being in the hyperplane $P = \{x \in \mathbb{R}^n : a_1 x_1 + \cdots + a_n x_n = b\}$ and satisfying $\|\hat{x}/u\| < 1 + \varepsilon$ is bounded by $2^{cn}$. Then there exists a randomized algorithm that solves BKP (2.2) with probability exponentially close to 1. The running time of the algorithm is $2^{O(n)}$.*

This theorem shows that if integral vectors of the hyperplane of BKP spread out *sparsely*, then we can hope to solve BKP "*efficiently*". Furthermore, this theorem also implies that BKP can be solved in time $2^{O(n)}$ if the number of solutions to it is not very large, e.g. $2^{cn}$. This is also the case for BIP.

COROLLARY 4.6. *Assume that there exist absolute constant $c, \varepsilon$ such that the number of integral vectors $\hat{x}$ being in $P = \{x \in \mathbb{R}^n : Ax = b\}$ and satisfying $\|\hat{x}/u\| < 1 + \varepsilon$ is bounded by $2^{cn}$. Then there exists a randomized algorithm that solves BIP (2.1) with probability exponentially close to 1. The running time of the algorithm is $2^{O(n)}$.*

*Proof.* [Proof of Theorem 4.5]

If we chose the parameters $s_0, s_1$, and $\delta$ large enough, the resulting lattice $L_0$ of the reduction in section 2 has the following properties.

For each lattice vector $y \in L_0\backslash S$ of $\ell_\infty$ length at most $1+\varepsilon$, there exists unique integral vector $\hat{x} \in P$ such that $\|\hat{x}/u\| < 1+\varepsilon$ and either $\hat{x}/u$ or $-\hat{x}/u$ is the head part of $y$. This property is easily derived from the observations in the proof of Lemma 2.11. Conversely, for each vector $\hat{x} \in P$ satisfying $\|\hat{x}/u\| < 1+\varepsilon$, there are at most two lattice vectors in $L_0\backslash S$ of $\ell_\infty$ length at most $1+\varepsilon$.[5] Moreover, the number of such vectors is bounded by $2^{cn}$ due to the hypothesis. Thus, the number of lattice vectors in $L_0\backslash S$ of $\ell_\infty$ length at most $1+\varepsilon$ is bounded by $2^{cn+1}$. Remember that, by Lemma 2.11, the shortest vector in $L_0\backslash S$ has $\ell_\infty$ length at most 1. This means the number of lattice vectors in $L_0\backslash S$ of length at most $(1+\varepsilon)\lambda_S(L_0)$ is bounded by $2^{cn+1}$, where $\lambda_S(L_0)$ is the length of the shortest vector in $L_0\backslash S$ for $\ell_\infty$ norm. From these observations, we may find the shortest vector in $L_0\backslash S$ in probabilistic time $2^{O(n)}$ (by Theorem 4.1). That is, BKP is solvable in probabilistic time $2^{O(n)}$. □

---

[5] One is $\left(\frac{\hat{x}}{u}, 0, e - \frac{\hat{x}}{u}, 0\right)$ and another is $-\left(\frac{\hat{x}}{u}, 0, e - \frac{\hat{x}}{u}, 0\right)$.

**Conclusion.** We obtain the new time bound for BIP; nonetheless, the time bound is still large. We believe that it can be substantially improved in both deterministic and probabilistic sense. An open question is *whether there is a randomized algorithm for BIP running in time $2^{O(n)}$?* Another one is *whether there is a deterministic algorithm for BIP running in time $2^{O(n)}$?* It seems that the first question is easier than the latter due to the fact that many lattice problems have this time bound, and that our reduction is an evidence.

**Acknowledgment.** The author would like to sincerely thank Daniele Micciancio for sharing his result in [8], and for many helpful discussions on lattice problems. Thanks are also to Jin-Yi Cai for pointing me to the reference [2].

**Appendix.** The following lemma is an observation in number theory and was proven in [6].

LEMMA 4.7. *Let $n, \lambda, t$ be integers such that $n$ is not a constant, $|\lambda| \geq \Theta(n), \gamma = 1 + \lambda + \cdots + \lambda^n$. Then the Diophantine equation*

$$(4.6) \qquad x_1 + \lambda x_2 + \cdots + \lambda^n x_{n+1} = \gamma.t$$

*has a solution of the form $(t, t, ..., t)$, and every other integral solution must have at least one component whose absolute value is not less than $\Theta(\lambda)$.*

*Proof.* We can rewrite equation (4.6) as follow:

$$(4.7) \qquad (x_1 - t) + \lambda(x_2 - t) + \cdots + \lambda^n(x_{n+1} - t) = 0$$

All integral solutions to (4.7) satisfy:

$$x_1 - t = \lambda.t_1$$

$$(4.8) \qquad (x_2 - t) + \lambda(x_3 - t) + \cdots + \lambda^{n-1}(x_{n+1} - t) = -t_1$$

where $t_1$ is an integer.

Similarly, all integral solutions to (4.8) satisfy:

$$x_2 - t = -t_1 + \lambda.t_2$$

$$(4.9) \qquad (x_3 - t) + \lambda(x_4 - t) + \cdots + \lambda^{n-2}(x_{n+1} - t) = -t_2$$

where $t_2$ is an integer.

By induction, all integral solutions to (4.6) satisfy:

$$(4.10) \qquad \begin{aligned} x_1 - t &= \lambda.t_1 \\ x_2 - t &= -t_1 + \lambda.t_2 \\ x_3 - t &= -t_2 + \lambda.t_3 \\ &\cdots\cdots\cdots\cdots \\ x_n - t &= -t_{n-1} + \lambda.t_n \\ x_{n+1} - t &= -t_n \end{aligned}$$

where $t_1, ..., t_n$ are integers.

It is clear that if $t_1 = \cdots = t_n = 0$, then $(t, ..., t) \in \mathbb{Z}^{n+1}$ is a solution to (4.6).

Now we consider any other integral solution $x$ to (4.6). There exist integers $t_1, ..., t_n$ satisfying (4.10), and at least one $t_r$ satisfying $t_r \neq 0$.

Let $t_l$ be the first non-zero element in the sequence $t_1, ..., t_n$, and $t_r$ be the last one. To observe the desired property of $x$ easily, we should examine the following cases:

*Case 1: $l > 1$*

With this assumption, we immediately have $x_1 = t, x_l = t + \lambda.t_l$. It is easy to see that if $|x_1| < \Theta(\lambda)$ then $x_l = \Theta(\lambda.t_l)$. If $|x_l| < \Theta(\lambda)$ then $t = \Theta(\lambda.t_l)$; consequently, $x_1 = \Theta(\lambda.t_l)$. In both cases, our claim is true.

*Case 2: $r < n$*

This case can be proved by the same argument as the one in *Case 1*.

*Case 3: $l = 1, r = n$* $\qquad (t_1 \neq 0, t_n \neq 0)$

It is easy to see that either $t = 0$, which yields $x_1 = \Theta(\lambda t_1)$, or $\exists |x_i| \geq \Theta(\lambda)$ will lead to the desired property of $x$.

Assume oppositely that $|t| \geq \Theta(\lambda) > 0$ and $|x_i| < \Theta(\lambda), \forall i \in [n+1]$. From (4.10), $|x_{n+1}| < \Theta(\lambda)$ implies $t_n = t - x_{n+1} = \Theta(t)$, $|x_n| < \Theta(\lambda)$ implies $t_{n-1} = \Theta(t + \lambda t_n) = \Theta(\lambda t)$ (due to $t_n \neq 0$),..., $|x_2| < \Theta(\lambda)$ implies $t_1 = \Theta(t + \lambda t_2) = \Theta(\lambda^{n-1} t)$. By the assumption $|x_1| < \Theta(\lambda)$, we have $t = \Theta(\lambda t_1) = \Theta(\lambda^n t)$. A contrary to the assumption $|t| \geq \Theta(\lambda)$ and $|\lambda| \geq \Theta(n)$ happens (due to the fact that $n$ is not a constant and neither is $\lambda$).

Assume oppositely that $|t| < \Theta(\lambda)$, $t \neq 0$ and $|x_i| < \Theta(\lambda), \forall i \in [n+1]$. From (4.10), $|x_n| < \Theta(\lambda)$ implies $|t_{n-1}| = |t - x_n + \lambda t_n| \geq \Theta(\lambda)$ (due to $t_n \neq 0$),..., $|x_2| < \Theta(\lambda)$ implies $|t_1| = |t - x_2 + \lambda t_2| \geq \Theta(\lambda^{n-1})$. By the assumption $|x_1| < \Theta(\lambda)$, we have $|t| = |x_1 - \lambda t_1| = \Theta(\lambda t_1) \geq \Theta(\lambda^n)$. A contrary to the assumption $|t| < \Theta(\lambda)$ and $|\lambda| \geq \Theta(n)$ happens. Consequently, the lemma follows. $\square$

**Disc.1:** solving CVP for $\ell_1$ norm

Let a target vector $t$ and a basis $H = (h_1, ..., h_m)$ of a lattice $L$ be the input of CVP. Kannan's algorithm [5] for CVP (for $\ell_2$ norm) is as follows. One chooses a suitable real $r$ (Kannan [5] finds a KZ-reduced basis of $L$ before choosing $r$) and considers the ball $B(t, r)$, for which $t$ and $r$ are the center and the radius, respectively. One can enumerate all lattice points in $L \cap B(t, r)$, and then choose the point $\hat{v}$ closest to $t$ as the output. It is not hard to see that if $B(t, r)$ contains $\hat{v} \in L$ closest to $t$ for $\ell_2$ norm, then the ball $B(t, 2r)$ must contain $\hat{u} \in L$ closest to $t$ for $\ell_1$ norm. Indeed, by the relation among norms, we have $\|t - \hat{v}\|_2 \leq \|t - \hat{u}\|_2 \leq \|t - \hat{u}\|_1 \leq \|t - \hat{v}\|_1 \leq \sqrt{2} \|t - \hat{v}\|_2$. Thus, with the ball $B(t, 2r)$, the enumeration step surely enumerates the vector $\hat{u} \in L$ closest to $t$ for $\ell_1$ norm. This means that we can adapt Kannan's algorithm to solve CVP for $\ell_1$ norm, with the caution that the bound should be doubled ($2A$ instead of $A$ as chosen in [3]). Using the analysis of Hanrot and Stehlé [3], we conclude that CVP for $\ell_1$ norm is solvable in deterministic time $poly(\psi) \cdot 2^{O(n)} \cdot n^{n/2 + o(n)} = poly(\psi) \cdot n^{n/2 + o(n)}$.

**Disc.2:** the assumption $\sum_{j=1}^{n} a_j u_j \neq 2b$

Assume now that the original BKP (2.2) satisfying $\sum_{j=1}^{n} a_j u_j = 2b$. Then we reduce it to the new one satisfying our desire. The new BKP is as follows:

*Find a vector $x \in \mathbb{Z}^{n+1}$ satisfying*

(4.11)
$$\begin{cases} a'_1 x_1 + \cdots + a'_{n+1} x_{n+1} = b' \\ 0 \leq x \leq u \\ 0 \leq x_{n+1} \leq 1 \end{cases}$$

Where $a'_j = a_j$, for all $j \leq n, a'_{n+1} = u_{\max}(n+1)b, b' = b + u_{\max}(n+1)b$.

It is not hard to see that solving the new BKP (4.11) is equivalent to solving (2.2). Indeed, if $(x_1, ..., x_n, 1)$ is a solution to (4.11), then $a_1 x_1 + \cdots + a_n x_n = b$. This implies $(x_1, ..., x_n)$ being a solution to (2.2). Moreover, $(x_1, ..., x_n, 0)$ cannot be a solution to (4.11) due to the fact that $a_1 x_1 + \cdots + a_n x_n = b'$ means a contrary to the assumption $a_j < b, \forall j \leq n$. These observations imply that $(x_1, ..., x_n, x_{n+1})$ is a solution to (4.11) if and only if $(x_1, ..., x_n)$ is a solution to (2.2).

Note that (4.11) satisfies $\sum_{j=1}^{n+1} a'_j u'_j \neq 2b'$, where $u'_j = u_j, \forall i \leq n$ and $u'_{n+1} = 1$. This means the new BKP satisfies our desired assumption.

## REFERENCES

[1] JOHANNES BLÖMER, *Closest vectors, successive minima, and dual hkz-bases of lattices*, in Proc. of the 27th Intern. Colloq. on Automata, Languages and Programming - ICALP, Springer-verlag, 2000, pp. 248–259.
[2] JOHANNES BLÖMER AND STEFANIE NAEWE, *Sampling methods for shortest vectors, closest vectors and successive minima*, in Proc. of the 34th Intern. Colloq. on Automata, Languages and Programming, Springer-verlag, 2007, pp. 65–77.
[3] GUILLAUME HANROT AND DAMIEN STEHLÉ, *Improved analysis of kannan's shortest lattice vector algorithm*, in Advanced in Cryptology - CRYPTO, Springer-verlag, 2007, pp. 170–186.
[4] RAVINDRAN KANNAN, *Polynomial-time aggregation of integer programming problems*, Journal of the ACM, 30 (1983), pp. 133–145.
[5] ———, *Minkowski's convex body theorem and integer programming*, Mathematics of Operations Research, 12 (1987), pp. 415–440.
[6] THÂN QUANG KHOÁT, *Reductions from knapsack problem to lattice problems*. Submitted, 2007.
[7] HENDRIK W. LENSTRA, *Integer programming with a fixed number of variables*, Mathematics of Operations Research, 8 (1983), pp. 538–548.
[8] DANIELE MICCIANCIO, *Efficient reductions among lattice problems*, in Proc. of the 19th annual ACM-SIAM Symposium on Discrete Algorithms -SODA, 2008, pp. 84–93.
[9] CHRISTOS H. PAPADIMITRIOU, *Computational Complexity*, Addison-Wesley, 1995.
[10] ALEXANDER SCHRIJVER, *Theory of Linear and Integer Programming*, John Wiley & Sons, 1999.